\documentclass{article}

\usepackage{arxiv}

\usepackage[utf8]{inputenc} 
\usepackage[T1]{fontenc}    
\usepackage{hyperref}       
\usepackage{url}            
\usepackage{booktabs}       
\usepackage{amsfonts,amsmath}       
\DeclareMathOperator{\Tr}{Tr}
\usepackage{nicefrac}       
\usepackage{microtype}      
\usepackage{lipsum}		
\usepackage{graphicx}
\usepackage{natbib}
\usepackage{doi}
\usepackage{caption}
\usepackage{subcaption}
\usepackage{float}

\title{Nonlinear analysis of a fiber-reinforced tubular conducting polymer-based soft actuator}

\author{ \href{https://orcid.org/0000-0002-1562-3687}{\includegraphics[scale=0.06]{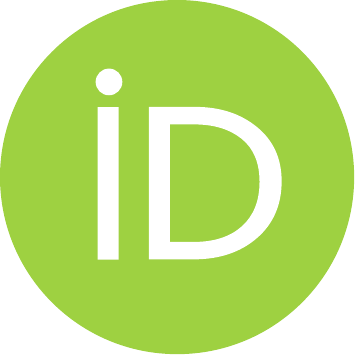}\hspace{1mm}Saswath Ghosh} \\
	Department of Applied Mechanics\\
	Indian Institute of Technology Delhi, \\ New Delhi, India, 110016. \\
	\texttt{S\_Ghosh.am.iitd.ac.in} \\
	\And
	\href{https://orcid.org/0000-0003-2720-4620}{\includegraphics[scale=0.06]{orcid.pdf}\hspace{1mm}Sitikantha Roy}\thanks{Corresponding author} \\
	Department of Applied Mechanics\\
	Indian Institute of Technology Delhi, \\ New Delhi, India, 110016. \\
	\texttt{sroy@am.iitd.ac.in} \\
}

\date{}

\renewcommand{\shorttitle}{FTCP actuator}

\begin{document}
\maketitle

\begin{abstract}
	This study presents the analytical modeling of a fiber-reinforced tubular conducting polymer (FTCP) actuator. The FTCP actuator is a low voltage-driven electroactive polymer arranged in an electrochemical cell. The electrochemical model is developed following an electrical circuit analogy that predicts the charge diffused inside the actuator for an applied voltage. An empirical relation is applied to couple the two internal phenomena, viz., diffusion of the ions and mechanical deformation. Further, the finite deformation theory is applied to predict the blocked force and free strain of the FTCP actuator. The developed model is consistent with existing experimental results for an applied voltage. In addition, the effect of various electrical and geometrical parameters on the performance of the actuator is addressed.
\end{abstract}


\section{Introduction}
In the past decades, research on electroactive polymers (EAPs)-based soft actuators has gained attention due to their wide engineering applications like an artificial muscle in exoskeletons, drug delivery systems, cell biology, and microactuators \citep{bar2004electroactive,hu2019pedot,farajollahi2016self}. Conducting polymers (CPs), also known as conjugated polymers, are promising materials for building a low voltage-driven (typically $1-3$ V) EAP-based soft actuator. The actuation mechanism of some standard CP actuators, made up of polypyrrole (PPy), polyaniline (PANi), and poly(3,4-ethylenedioxythiophene) (PEDOT), have been studied in \citep{hu2019pedot, madden2000conducting,kaneto2014conducting}. These behave as a semiconductor in the neutral state and become conductive upon chemical/electrochemical oxidation. The oxidation generates polarons (unit of positive charges) that attract opposite charged mobile ions. The electrolyte contains the anions and cations, which get diffused in/out accordingly into the polymer resulting in volume change.

In general, CP actuators can be classified as anion or cation-driven, which can perform different actuation modes like linear (axial), torsion, and bending deformation \citep{hu2019pedot,melling2019conjugated,fang2010fiber}. One of the primitive models for free-standing CP films exhibiting linear deformation was developed by \citep{madden2000conducting}. In the diffusive-elastic-metal model, the electrochemical process in the CP actuator is represented by an electrical circuit. The equivalent electrical circuit is solved to find the total admittance, hence the current and charge density stored in the polymer. Finally, the charge density is empirically related to the volumetric strain induced in the polymer due to the diffusion of ions. However, linear deformation theory was used to find the strain and stress of the actuator \citep{madden2000conducting}. Later, the same approach was utilized by \citep{fang2008nonlinear} to develop a finite deformation theory-based trilayer bending actuator. Likewise, a few tubular conducting polymer actuators exhibiting axial deformation \citep{ding2003high,samani2004mechanical,yamato2006tubular}, torsional \citep{fang2010fiber}, and bending deformation \citep{farajollahi2016self} have been designed. In \citep{ding2003high}, experimental characterization of a helical wire wrapped around the PPy tube for braille application was performed. A viscoelastic model of a helical wire wrapped around the PPy tube for braille application has been developed in \citep{samani2004mechanical}. Moreover, the wire wrapped around the cylinder was to increase the electrical conductivity. They \citep{ding2003high,samani2004mechanical} did not study effect of wire wrapping on the deformation of the actuator. Similarly, in \citep{yamato2006tubular} a tubular CP actuator without wire wrapping is designed to perform axial actuation. Further, in \citep{fang2010fiber}, a fiber-reinforced torsional CP actuator following nonlinear elastic theory was developed. The model highlights the need for nonlinear deformation theory for CP actuators \citep{fang2010fiber,fang2008nonlinear,sendai2009anisotropic}. To the best of the authors' knowledge, none of the models describes the behavior of an axially deforming fiber-reinforced tubular conducting polymer (FTCP) actuator. In this work, the FTCP actuator consists of two fiber families coiled around the tube axis, exhibiting an axial deformation is analyzed using the finite deformation theory. Herein, the diffusion equation is solved in polar coordinates ($R, \Theta, Z$) to estimate the steady-state charge stored in the actuator. It is significant to mention that the fiber properties control the axial elongation or contraction of the actuator for the same applied voltage \citep{demirkoparan2015magic}, discussed in detail in Section \ref{section-3}. The developed actuator can be combined to design soft locomotive robots similar to \citep{shepherd2011multigait,calisti2017fundamentals}.

In the present work, a physics-based electro-chemo-mechanical (ECM) deformation model is formulated to predict the response of a fiber-reinforced tubular conducting polymer (FTCP) actuator. The electrochemical process is modeled utilizing the electrical circuit analogy approach to obtain the admittance expression in cylindrical space coordinates. The formulated expression is used to determine the volume change of the actuator as a result of ion diffusion. Further, a finite deformation theory is utilized to predict the response of the actuator for a given volume change in the polymer. The model predicts the blocked force and free expansion/contraction of the FTCP actuator for an applied voltage. Moreover, we compare the frequency response of the FTCP actuator with existing works qualitatively. Further, the effects of the electrical circuit and geometrical parameters on the actuation are discussed. 

The remaining sections are structured as follows: the ECM deformation model of the FTCP actuator is described in Section \ref{section-2}. In Section \ref{section-3}, the response of the actuator for an applied voltage is discussed. In addition, the effect of various model parameters has been addressed. Lastly, Section \ref{section-4} summarises the main inferences of the developed model. Table \ref{tab:1} provides a list of all variables for a quick overview.

\begin{table}[h]
\caption{List of symbols.}
\begin{tabular}{lll}
	\toprule
Symbols &Description (Units)\\
\midrule
$t$  &Total time (s)\\
$s$  &Laplace variable (1/s)\\
$v_{in} (t) \, [V_{in} (s)]$  &Applied step voltage in time [Laplace] domain (V)\\
$i (t) \, [I(s)]$  &Total circuit current in time [Laplace] domain (A)\\
$R_e$  &Equivalent circuit resistance ($\Omega$)\\
$C_{dl}$  &Double layer capacitance (F)\\
$i_c (t) \,[I_c(s)]$  &Double layer charging current in time [Laplace] domain (A)\\
$i_d (t)\, [I_d(s)]$ &Diffusion current in time [Laplace] domain (A)\\
$Q^{'}(s)$  &Total ionic charge in double layer (C)\\
$c(R,t)\,[C(R,s)]$  &Mobile ion concentration inside conducting polymer along radial direction\\ 
 &in time [Laplace] domain (mol/$\text{m}^3$)\\
$F$ &Faraday's constant (C/mol)\\
$A$ &Surface area of the double layer capacitor ($\text{m}^2$)\\
$\delta$ &Double layer thickness (m)\\
$D$ &Diffusion coefficient ($\text{m}^2$/s)\\
$\boldsymbol{j}$ &Ionic flux vector (mol/s-$\text{m}^2$)\\
$Y(s)$ &Circuit admittance (S)\\
$\nu$ &Volume ratio of FTCP actuator\\
$V_f$ &Deformed volume of FTCP actuator ($\text{m}^3$)\\
$V_0$ &Undeformed volume of FTCP actuator ($\text{m}^3$)\\
$\kappa$ &Volumetric coupling coefficient ($\text{m}^3$/C)\\
$Q$ &Total steady-state charge stored in FTCP per unit its undeformed volume (C/$\text{m}^3$)\\
$R_1$ &Undeformed inner radius of the actuator (m)\\
$R_2$ &Undeformed outer radius of the actuator (m)\\ 
$L$ &Undeformed length of the actuator (m)\\ 
$r_1$ &Deformed inner radius of the actuator (m)\\
$r_2$ &Deformed outer radius of the actuator (m)\\ 
$l$ &Deformed length of the actuator (m)\\ 
$\phi$ &Fiber angle  (degree)\\
$\beta$ &Shear modulus of the actuator (MPa)\\
$\zeta$  &Anisotropic factor representing the fiber strength\\
$F_{axial}$ &Axial force of FTCP actuator (N)\\ 
$\lambda_{3}$  &Axial stretch of FTCP actuator\\
\bottomrule
\end{tabular}
\label{tab:1}
\end{table}

\section{Model description}
\label{section-2}

\begin{figure}[h]
      \centering
      \includegraphics [trim=0.4cm 0.2cm 1.2cm 0.2cm, clip=true,width=0.7 \linewidth]{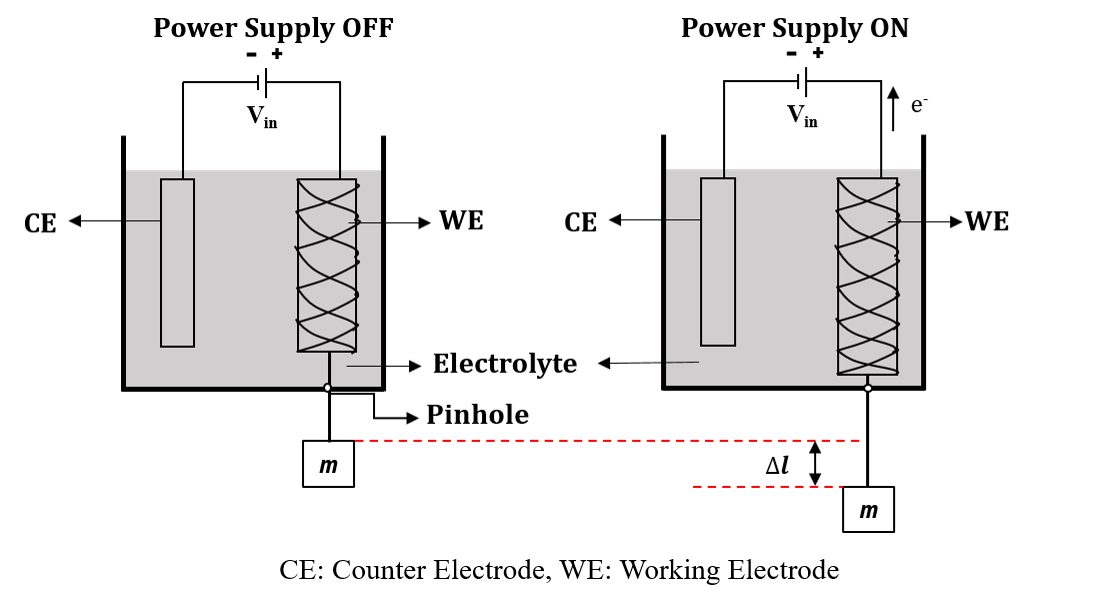}
      \caption{Schematic representation of an electrochemical cell consisting of a FTCP actuator submerged in an electrolyte.}
      \label{fig:1}
\end{figure}

Consider the working electrode of an electrochemical cell made of a fiber-reinforced tubular conducting polymer (FTCP) as shown in Figure \ref{fig:1}. The oppositely charged ions diffuse into the polymer as the potential is applied between the counter electrode and FTCP. The axial deformation in the actuator is modeled under the following assumptions:
\begin{enumerate}
\item No account of solvent diffusion with the ions has been considered.
\item The ECM coupling is purely empirical. In other words, the volume ratio is proportional to the charge diffused into the actuator, and the proportionality constant is determined from experiments.
\item The volume ratio is uniform and not varying along the radius of the polymer.
\end{enumerate}

\begin{figure}[h]
      \centering
      \includegraphics [trim=0cm 0.2cm 0cm 0cm, clip=true,width=1 \linewidth]{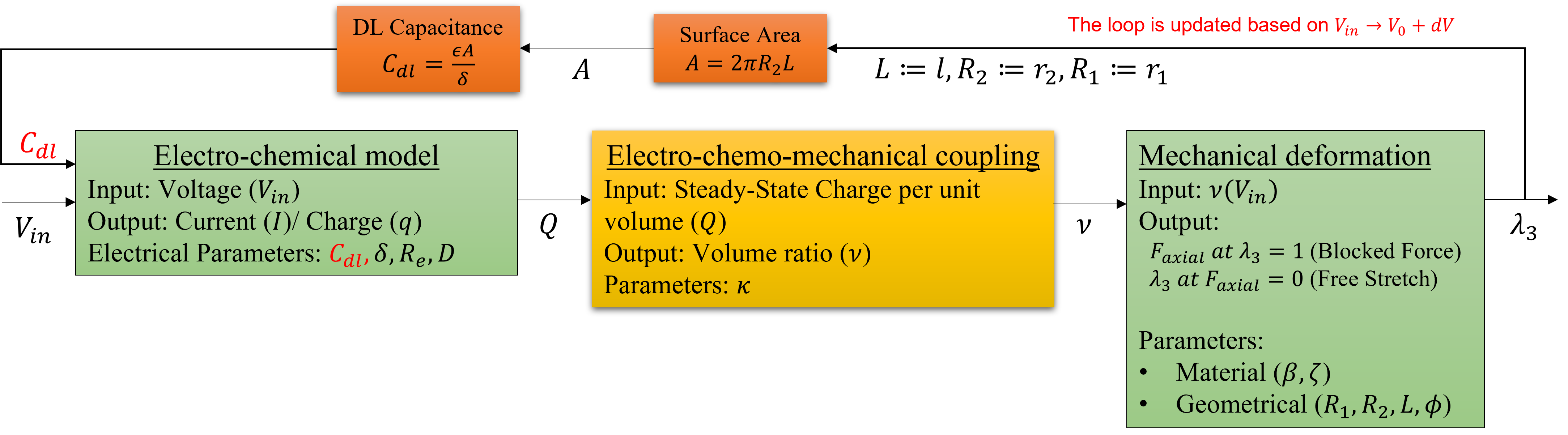}
      \caption{Block diagram representing the ECM deformation of FTCP actuator.}
      \label{fig:2}
\end{figure}

The model combines three sub-domains as shown in Figure \ref{fig:2}. The first block is an electrochemical model that determines the steady-state charge stored in the actuator for an applied voltage. The second block connects the stored charge to the volume change in the polymer. The last block relates the volume change to the axial deformation in the FTCP actuator. Since the relation between blocked force (or free stretch) and the applied voltage is nonlinear, the voltage can be applied to the actuator incrementally. Consequently, the change in surface area after every increment yields a change in the double layer capacitance of the actuator. Thus, the ECM model is solved using an updated capacitance value for each step voltage increment. Furthermore, a detailed discussion of each sub-domain follows.

\subsection{Electrochemical model}
\begin{figure}[h]
      \centering
      \includegraphics [trim=0cm 0.2cm 5cm 0cm, clip=true,width=0.7 \linewidth]{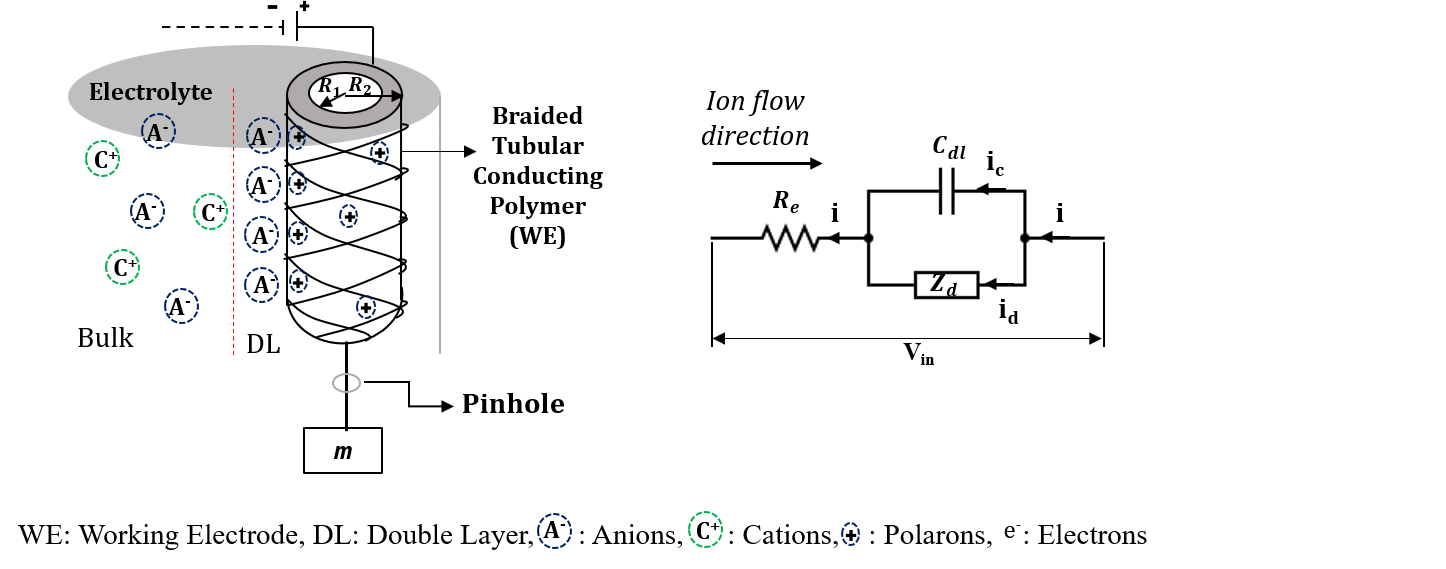}
      \caption{Equivalent circuit representing the ion transfer mechanism.}
      \label{fig:3}
\end{figure}

The electrochemical (EC) model relates the input voltage, and output current of the EC cell through an electrical circuit analogy as shown in Figure \ref{fig:3}.  The applied potential forms a double layer (DL) between FTCP and the surrounding electrolyte that contains oppositely charged ions. The ions in the DL get diffused into the FTCP due to concentration gradient. The total potential drop across the circuit is given by
\begin{equation}\label{eqn:1}
    v_{in}(t) = i(t)R_{e}+\frac{1}{C_{dl}}\int_{0}^{t}{i_c(t)}dt,
\end{equation}
where $i(t)$ is total circuit current, $i_c(t)$ is DL charging current, and $C_{dl}$ is DL capacitance. $R_e$ includes the resistance of ion movement in the electrolyte as well as electrical resistance at connecting ends. The total current in the circuit is given by
\begin{equation}\label{eqn:2}
    i(t) =i_c(t) +i_d(t),
\end{equation}
where $i_d(t)$ is diffusion current.

Equation \ref{eqn:1} and \ref{eqn:2} can be represented using Laplace variable, $s$, as
\begin{equation}\label{eqn:3}
\begin{split}
    V_{in}(s) &= I(s)R_e + \frac{I_c(s)}{sC_{dl}},\\
    I(s) &= I_c(s) + I_d(s).\\
\end{split}
\end{equation}
The DL charging current can be expressed in terms of mobile ion concentration as
\begin{equation}\label{eqn:4}
\begin{split}
    I_c(s)=sQ^{*}(s)=sC(R,s)FA\delta.\\
\end{split}
 \end{equation}
where $Q^{*}(s)$ is total charge of ions in DL at any time, $C(R,s)$ is the concentration of ions inside the polymer at any given time, $A$ is the capacitor surface area, and $\delta$ is the DL thickness. The Fick's first law of diffusion \citep{bird2006transport}, $\textbf{j} = -D \, \mathrm{Grad} \, \textbf{c}$,
where $\mathbf{j}$ is ionic flux vector through the surface normal or current density \citep{bard2001fundamentals}, $D$ is the diffusion coefficient, and $\textbf{c}$ is the ionic concentration per unit polymer volume at any time, is applied to find the diffusion current in the circuit.
The ionic diffusion current \citep{bard2001fundamentals} is expressed as 
\begin{equation}\label{eqn:6}
\begin{split}
    \mathbf{j} =\frac {i_d(t)}{nFA},\\ 
\end{split}
\end{equation}
where $n$ is the valency of the ion. Considering one-dimensional diffusion of univalent ions from outer to inner surface of FTCP electrode, the diffusion current is given by
\begin{equation}\label{eqn:7}
    I_d(s) = FAD\dfrac{\partial C(R,s)}{\partial R}.
\end{equation}
Now, to find the expression of $I(s)$ in terms of concentration of ions, we used Fick’s second law of diffusion \citep{bird2006transport}. The one-dimensional transport equation in cylindrical coordinates ($R, \theta, Z$) is expressed as
\begin{equation}\label{eqn:8}
\begin{split}
    \frac{\partial c(R,t)}{\partial t} = D\left(\frac{\partial ^2 c(R,t)}{\partial R^2} +\frac{1}{R} \frac{\partial c(R,t)}{\partial R} \right),
\end{split}
\end{equation}
where $c(R,t)$ is the concentration of ion along radial direction in time domain. Equation \ref{eqn:8} is solved analytically using Laplace transform method. The method converts partial differential equation into a second order ordinary differential equation. The Laplace transform of equation \ref{eqn:8} is 
\begin{equation}\label{eqn:9}
\begin{split}
    \mathcal{L}\left(\frac{\partial c}{\partial t}\right) = sC(R,s) -c(R,0),\\
    \mathcal{L} \left(\frac{\partial ^2 c}{\partial R^2}\right) = \frac{d^2C(R,s)}{dR^2}, \quad \mathcal{L} \left(\frac{\partial c}{\partial R}\right) = \frac{dC(R,s)}{dR}.\\
\end{split}
\end{equation}
The initial condition $c(R,0)=0$ is utilised. Substituting equation \ref{eqn:9} in equation \ref{eqn:8} and rearranging the terms, we have
\begin{equation}\label{eqn:10}
\begin{split}
    \frac{d^2 C(R,s)}{dR^2} + \frac{1}{R} \frac{dC(R,s)}{dR} -\frac{s}{D}C(R,s) = 0.
\end{split}
\end{equation}
The above second-order ordinary differential equation (\ref{eqn:10}) resembles the modified Bessel equation whose solution can be expressed in terms of modified Bessel functions. \citep{kreyszig2009advanced,Mathematica}. Thus, the concentration of ions diffused inside the polymer is expressed as 
\begin{equation}\label{eqn:11}
\begin{split}
    C(R,s) = \alpha_1 \mathrm{I_0} \left( \sqrt{\frac{s}{D}} R\right) +\alpha_2 \mathrm{K_0} \left( \sqrt{\frac{s}{D}} R\right),
\end{split}
\end{equation}
where
\begin{equation}\label{eqn:13}
\begin{split}
\alpha_1 = \Bigg[I_c(s)D \mathrm{K_1}\left(\sqrt{\frac{s}{D}} R_2\right) +I_d(s)\delta \sqrt{sD} \mathrm{K_0}\left(\sqrt{\frac{s}{D}} R_2\right)\Bigg]
\Bigg[sFAD\delta  \Bigg[\mathrm{I_1}\left(\sqrt{\frac{s}{D}} R_2\right)  \mathrm{K_0}\left(\sqrt{\frac{s}{D}} R_2\right)\\
+\mathrm{I_0}\left(\sqrt{\frac{s}{D}} R_2\right)\mathrm{K_1}\left(\sqrt{\frac{s}{D}} R_2\right)
 \Bigg]\Bigg]^{-1},\\
\alpha_2 = \Bigg[I_c(s)D \mathrm{I_1}\left(\sqrt{\frac{s}{D}} R_2\right) -I_d(s)\delta \sqrt{sD} \mathrm{I_0}\left(\sqrt{\frac{s}{D}} R_2\right)\Bigg]
\Bigg[sFAD\delta \Bigg[\mathrm{I_1}\left(\sqrt{\frac{s}{D}} R_2\right)\mathrm{K_0}\left(\sqrt{\frac{s}{D}} R_2\right) \\ +\mathrm{I_0}\left(\sqrt{\frac{s}{D}} R_2\right)\mathrm{K_1}\left(\sqrt{\frac{s}{D}} R_2\right)  \Bigg]\Bigg]^{-1},\\
\end{split}
\end{equation}

is obtained using two conditions $C(R=R_2,s) = {I_c(s)}/(sFA\delta)$, and ${\partial C(R=R_2,s)}/{\partial R} = {I_d(s)}/(FAD)$, at the interface between electrolyte and outer radius of FTCP. $\mathrm{I_0}\left(\sqrt{{s}/{D}} R_2\right)$, and $\mathrm{I_1} \left(\sqrt{{s}/{D}} R_2\right)$ are the modified Bessel functions of first kind of order zero and one, respectively. $\mathrm{K_0} \left(\sqrt{{s}/{D}} R_2\right), \mathrm{K_1} \left(\sqrt{{s}/{D}} R_2\right)$ are the modified Bessel functions of second kind of order zero and one, respectively.
Rewriting equation \ref{eqn:3}, we have
\begin{equation}\label{eqn:14}
\begin{split}
    I_c(s) &= sC_{dl}(V_{in} -I(s)R_e),\\
    I_d(s) &= I(s)-I_c(s) = I(s)-sC_{dl}(V_{in} -I(s)R_e).\\
\end{split}
\end{equation}
Substituting equation \ref{eqn:14} into equation \ref{eqn:13} and assuming the inner boundary is sealed and no diffusion is taking place, i.e., ${\partial C(R=R_1,s)}/{\partial R} =0$, we have
\begin{align}\label{eqn:15}
\begin{split}
I(s) \Bigg[\bigg((1+sC_{dl}R_e)\delta s\mathrm{K_0}\left(\sqrt{\frac{s}{D}} R_2\right) - sC_{dl}R_e\sqrt{sD} \mathrm{K_1}\left(\sqrt{\frac{s}{D}} R_2\right)\bigg)\mathrm{I_1}\left(\sqrt{\frac{s}{D}} R_1\right)\\+ \bigg((1+sC_{dl}R_e)\delta s \mathrm{I_0}\left(\sqrt{\frac{s}{D}} R_2\right) + sC_{dl}R_e\sqrt{sD}\mathrm{I_1}\left(\sqrt{\frac{s}{D}} R_2\right)\bigg) \mathrm{K_1}\left(\sqrt{\frac{s}{D}} R_1\right)\Bigg]\Bigg]\\
= V_{in}(s)\Bigg[\bigg(sC_{dl}R_e\sqrt{sD} \mathrm{I_1}\left(\sqrt{\frac{s}{D}} R_2\right)- sC_{dl}\sqrt{sD}\mathrm{K_1}\left(\sqrt{\frac{s}{D}} R_2\right)\bigg)\mathrm{I_1}\left(\sqrt{\frac{s}{D}} R_1\right) \\+\left( sC_{dl}\delta s \times \mathrm{I_0}\left(\sqrt{\frac{s}{D}} R_2\right) + sC_{dl}R_e\sqrt{sD}\mathrm{I_1}\left(\sqrt{\frac{s}{D}} R_2\right)\right)\mathrm{K_1}\left(\sqrt{\frac{s}{D}} R_1\right)\Bigg]\Bigg].\\ 
\end{split}
\end{align}

Further, the expression is simplified and the admittance of the circuit is given by 

\begin{equation}\label{eqn:16}
\begin{split}
Y(s) = \frac{I(s)}{V_{in}(s)} 
 = \frac{s\delta +\sqrt{sD} \frac{\left(\mathrm{I_{12}} \mathrm{K_{11}} - \mathrm{I_{11}}\mathrm{K_{12}}\right)}{\left(\mathrm{I_{11}} \mathrm{K_{02}} + \mathrm{I_{02}} \mathrm{K_{11}}\right)}}{\left(\frac{\delta}{C_{dl}} +s\delta R_e \right)  +\sqrt{sD}R_e \frac{\left(\mathrm{I_{12}}\mathrm{K_{11}} - \mathrm{I_{11}}\mathrm{K_{12}}\right)}{\left(\mathrm{I_{11}}\mathrm{K_{02}}  +\mathrm{I_{02}}\mathrm{K_{11}}\right)}},
\end{split}
\end{equation}
where $\mathrm{I_{mn}}$ is shorthand notation for modified Bessel function of first kind of order $m(=0,1)$ and function argument as $R_n \sqrt{{s}/{D}}(n=1,2)$ for inner and outer radius of the cylinder, respectively. Similarly, $\mathrm{K_{mn}}$ is shorthand notation for modified Bessel function of second kind of order $m(=0,1)$ and function argument as $R_n \sqrt{{s}/{D}} (n=1,2)$ for inner and outer radius of the cylinder, respectively. For example, $\mathrm{I_{02}}$ represents $\mathrm{I_0}\left(R_2 \sqrt{{s}/{D}}\right)$. The total current in the circuit is determined for a given voltage increment. Moreover, the total charge stored in the polymer can be determined by integrating the total current over time period. The circuit parameters $R_e, C_{dl}, D, \delta$ needs to be estimated from experiments for a given electrolyte and FTCP material. 

\subsection{Electro-chemo-mechanical (ECM) coupling}
Assuming that the ions diffused into polymer is the only reason for change in the polymer volume. The ratio of final volume, $V_f$, to initial volume, $V_0$, of the FTCP actuator, $\nu$ is given by 
\begin{equation}\label{eqn:17}
\begin{split}
    \nu = \frac{V_f}{V_0} = 1+ \kappa Q,
\end{split}    
\end{equation}
where $\kappa$ is the coupling coefficient that is to be determined using experiments. $Q$ is the total steady-state charge induced in the polymer per unit undeformed volume. It is determined from the above EC model (\ref{eqn:16}) using final value theorem when $s \longrightarrow 0$. It can be expressed as $Q = q/V_0$, where 
\begin{align}\label{eqn:117}
    q = C_{dl}V_{in}\left(1+\frac{R_2}{2\delta}\left(1-\left(\frac{R_1}{R_2}\right)^2\right)\right).
\end{align}
Herein, the volume ratio can be measured using polymer dimensions. The total charge consumed by the FTCP can be measured using coulovoltametric response plots \citep{otero2014ionic}. Eventually, $\kappa$ is estimated through curve fitting technique. Further, the volume ratio is related to the axial stretch of the CP actuator. Although the nonlinear volume charge relation is suggested for CP films at higher external stress ($>2$ MPa) \citep{sendai2009anisotropic}. The influence of external load on actuator deformation will be the subject of future research.  

\subsection{Mechanical deformation model}
\begin{figure}[h]
      \centering
      \includegraphics [trim=0cm 0cm 0.2cm 0cm, clip=true,width=0.25 \linewidth]{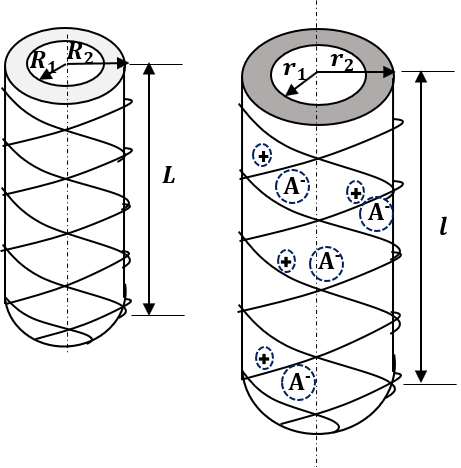}
      \caption{Undeformed and deformed configuration of a conducting polymer (CP) actuator.}
      \label{fig:4}
\end{figure}

Consider a fiber-reinforced tubular conducting polymer (FTCP) actuator as shown in Figure \ref{fig:4}. 
The undeformed reference configuration $\mathcal{B}_0$ is described in terms of cylindrical coordinates $(R,\Theta, Z)$ as
\begin{align}\label{eqn:18}
	\begin{split}
		R_1 \leq R \leq R_2, \quad 0\leq \Theta \leq 2 \pi, \quad 0\leq Z \leq L,
	\end{split}
\end{align}
where $R_1$, $R_2$ and $L$ represent the undeformed inner radius, outer radius and length of the cylindrical actuator, respectively. Similarly, the deformed configuration $\mathcal{B}$ is described in terms of the coordinates $(r,\theta, z)$ by 
\begin{align}\label{eqn:19}
	\begin{split}
		r_1 \leq r \leq r_2, \quad 0\leq \theta \leq 2 \pi, \quad 0\leq z \leq l,
	\end{split}
\end{align}
where $r_1$, $r_2$ and $l$ are deformed inner radius, outer radius and length of the actuator, respectively. The deformation map for an inflation and extension in the cylinder is given by
\begin{equation}\label{eqn:20}
    r=r(R), \quad \theta=\Theta, \quad z=\lambda_3 Z,
\end{equation}
where $\lambda_3=\frac{l}{L}$ is the axial stretch in the actuator. For the given deformation map, the deformation gradient tensor \textbf{F} is defined as
\begin{align}\label{eqn:21}
\mathbf{F} = \lambda_1 \mathbf{e_R}\otimes\mathbf{e_r} + \lambda_2 \mathbf{e}_{\Theta} \otimes\mathbf{e_\theta}+ \lambda_3 \mathbf{e_Z}\otimes\mathbf{e_z},
\end{align}
where $\lambda_1=\frac{dr}{dR}$, $\lambda_2=\frac{r}{R}$ and $\lambda_3=\frac{l}{L}$ represent the corresponding principal stretches of the actuator. The fiber imposed in one direction is described in the undeformed configuration as
\begin{equation}\label{eqn:22}
\mathbf{M} = (\sin \phi) \mathbf{e_\Theta} + (\cos \phi)\mathbf{e_Z},
\end{equation}
where $\phi$ is the fiber angle. The fiber imposed in another direction is described in the undeformed configuration as
\begin{equation}\label{eqn:23}
\mathbf{M'} = -(\sin \phi) \mathbf{e_\Theta} + (\cos \phi) \mathbf{e_Z}.
\end{equation}
The two fibers are used to wrap the cylindrical conducting polymer and making it geometrically symmetric. The set of invariant for an anisotropic hyperelastic material is given by 
\begin{equation}\label{eqn:24}
\begin{split}
I_1 &= \Tr \mathbf{B} = \lambda_1^2 + \lambda_2^2 +\lambda_3^2,\\
I_2 &= \frac{1}{2} [(\Tr\mathbf{B})^2 - \Tr \mathbf{B}^2] = {\lambda_2^2}{\lambda_3^2} + {\lambda_1^2}{\lambda_3^2} +{\lambda_1^2}{\lambda_2^2},\\
I_3 &= \det\mathbf{B} = (\lambda_1\lambda_2\lambda_3)^2 =\nu^2,\\
I_4 &= \mathbf{m}\cdot\mathbf{m} = \mathbf{FM}\cdot\mathbf{FM} = \lambda_2^2 \sin^2 \phi + \lambda_3^2 \cos^2 \phi,\\
I_5 &= \mathbf{m}\cdot\mathbf{Bm} =\lambda_2^4 \sin^2 \phi + \lambda_3^4 \cos^2 \phi,\\
I_6 &= \mathbf{m'}\cdot\mathbf{m'} = \mathbf{FM'}\cdot\mathbf{FM'} = \lambda_2^2 \sin^2 \phi + \lambda_3^2 \cos^2 \phi = I_4,\\
I_7 &= \mathbf{m'}\cdot\mathbf{Bm'} =\lambda_2^4 \sin^2 \phi + \lambda_3^4 \cos^2 \phi,\\
I_8 &= \mathbf{m}\cdot\mathbf{m'} =-\lambda_2^2 \sin^2 \phi + \lambda_3^2 \cos^2 \phi,
\end{split}
\end{equation}
where $\mathbf{B=FF^T}$ is the left Cauchy-Green deformation tensor. $\mathbf{m=FM}$, and $\mathbf{m'=FM'}$ are the corresponding fiber direction in the deformed configuration.

In this work, the modified Neo-Hookean model for matrix and standard fiber model for fibers \citep{treloar1975physics,demirkoparan2007swelling,merodio2002material} have been employed. The total strain energy density function is the summation of strain energy density function of matrix ($W_m$) and fibers ($W_f)$ \citep{merodio2002material}, and is given as
\begin{equation}\label{eqn:25}
\begin{split}
    W &= W_m + W_f,\\
    W &= \frac{1}{2} \beta [(I_1 - 3\nu^{2/3})+ \zeta \left((I_4-1)^2 + (I_6-1)^2\right)],
\end{split}
\end{equation}
where $\beta$ is the material constant and $\zeta$ is the anisotropic parameter describing the fiber strength \citep{merodio2002material,feng2013measurements} to be determined through experiments. The Cauchy stress is defined as 
\begin{equation}\label{eqn:26}
\begin{split}
    \mathbf{\sigma} = -p\mathbf{I} + \frac{2}{\nu}\frac{\partial W}{\partial I_1}\mathbf{B} + \frac{2}{\nu}\frac{\partial W}{\partial I_4}\mathbf{m}\otimes\mathbf{m} + \frac{2}{\nu}\frac{\partial W}{\partial I_6}\mathbf{m'}\otimes\mathbf{m'}
\end{split}
\end{equation}
where $p$ is the Lagrange multiplier. The principal stress components are expressed as
\begin{equation}\label{eqn:27}
\begin{split}
    \sigma_{rr} &= -p + \frac{\beta}{\nu}\lambda_1^2,\\
    \sigma_{\theta \theta} &= -p + \frac{\beta}{\nu}\lambda_2^2 + \frac{4 \beta \zeta}{\nu} (I_4 -1)\left(\lambda_2^2 \sin^2 \phi\right),\\
    \sigma_{zz} &= -p + \frac{\beta}{\nu}\lambda_3^2 + \frac{4 \beta \zeta}{\nu} (I_4 -1)\left(\lambda_3^2 \cos^2 \phi\right).\\
\end{split}
\end{equation}
Assuming that the ions are uniformly distributed over the volume, i.e., $\nu \neq \nu(R)$ and integrating $I_3$ to get relation between undeformed and deformed radius, axial stretch, and volume ratio, we have
\begin{equation}\label{eqn:28}
\begin{split}
    r^2 = r_1^2 + \left(\frac{\nu}{\lambda_3}\right)(R^2-R_1^2).
\end{split}
\end{equation}
An additional condition is assumed, i.e., the invariant $I_4$ is independent of $R$. Thus, we have  $r_1^2 = \left({\nu}/{\lambda_3}\right)R_1^2$. Essentially, it implies $\lambda_1 = \lambda_2 = \sqrt{\nu/\lambda_3}$. 
The equilibrium equation in absence of any body force in
cylindrical coordinates is given by 
\begin{equation}\label{eqn:31}
\begin{split}
    \frac{d\sigma_{rr}}{dr} + \frac{1}{r}(\sigma_{rr} -\sigma_{\theta \theta}) =0.
\end{split}
\end{equation}
Solving the above equilibrium equation \ref{eqn:31} and applying the boundary condition $\sigma_{rr}(r=r_1) =0$, we have
\begin{equation}\label{eqn:32}
\begin{split}
    \sigma_{rr}(r) = \frac{4 \beta \zeta}{\lambda_3}\sin^2 \phi \left(\frac{\nu}{\lambda_3} \sin^2 \phi +\lambda_3^2 \cos^2 \phi -1 \right) \ln\left(\frac{r}{r_1}\right),\\
    \sigma_{\theta \theta}(r) = \sigma_{rr}(r) +  \frac{4\beta \zeta}{\lambda_3}  \sin^2 \phi \bigg(\frac{\nu}{\lambda_3} \sin^2 \phi +\lambda_3^2 \cos^2 \phi -1 \bigg),\\
    \sigma_{zz}(r) = \sigma_{rr}(r) + \frac{\beta}{\nu} \bigg(\lambda_3^2 -\frac{\nu}{\lambda_3} +4 \zeta \lambda_3^2 \cos^2 \phi  \left(\frac{\nu}{\lambda_3} \sin^2 \phi +\lambda_3^2 \cos^2 \phi -1 \right) \bigg).\\
\end{split}
\end{equation}
The axial force output of the actuator is expressed as
\begin{align}\label{eqn:33}
    F_{axial} = 2 \pi \int_{r_1}^{r_2} \sigma_{zz} rdr.
\end{align}
Finally, we obtain the output force as  
\begin{equation}\label{eqn:34}
\begin{split}
    F_{axial} = 2 \pi \Bigg[\frac{4\beta \zeta}{ \lambda_3}\sin^2 \phi \left(\frac{\nu}{\lambda_3} \sin^2 \phi +\lambda_3^2 \cos^2 \phi -1 \right) \left(\frac{r_2^2}{2} \ln{\frac{r_2}{r_1}} -\frac{r_2^2 -r_1^2}{4} \right)\\
    + \frac{\beta}{\nu} \left(\frac{r_2^2 -r_1^2}{2}\right) \bigg(\lambda_3^2 -\frac{\nu}{\lambda_3} +4 \zeta \lambda_3^2 \cos^2 \phi \bigg(\frac{\nu}{\lambda_3} \sin^2 \phi +\lambda_3^2 \cos^2 \phi -1 \bigg) \bigg)\Bigg].\\
\end{split}
\end{equation}
The blocked force can be obtained from above equation \ref{eqn:34} at no stretch ($\lambda_3 = 1$) condition. Similarly, the free stretch of the actuator for an applied voltage can be obtained by setting the output force expression (\ref{eqn:34}) to zero. The nonlinear equation is given by
\begin{align}
\begin{split}\label{eqn:37}
    \Bigg [\frac{4 \zeta}{ \lambda_3}\sin^2 \phi \left(\frac{\nu}{\lambda_3} \sin^2 \phi +\lambda_3^2 \cos^2 \phi -1 \right) \bigg(\frac{r_2^2}{2} \ln{\frac{r_2}{r_1}} -\frac{r_2^2 -r_1^2}{4} \bigg)
     + \left(\frac{r_2^2 -r_1^2}{2\nu}\right)
     \bigg(\lambda_3^2 -\frac{\nu}{\lambda_3} \\+4 \zeta \lambda_3^2 \cos^2 \phi \left(\frac{\nu}{\lambda_3} \sin^2 \phi +\lambda_3^2 \cos^2 \phi -1 \right) \bigg) \Bigg]=0.\\
\end{split}
\end{align}
The expression is solved using Newton-Raphson method in MATLAB.

\section{Results and discussions}\label{section-3}
In this section, we analyze the behavior of the fiber-reinforced tubular conducting polymer (FTCP) actuator for different applied step voltages. The values of the electrical circuit, geometrical, and material parameters used in the simulation are summarised in Table \ref{tab:2}.

\begin{table}[H]
	\caption{Model parameters used in the simulation}
	\centering
	\begin{tabular}{lll}
		\toprule
		Parameter     & Value  \\
		\midrule
$L$\cite{ding2003high} & $30$ mm\\
$R_1$\cite{ding2003high} & $125$ $\mu$m\\
$R_2$\cite{ding2003high} & $200$ $\mu$m\\
$\phi$& $15^{\circ}$\\
$\beta$ \cite{kaneto2014conducting} &$280$ MPa \\
$\zeta$ &$0.5$\\
$R_e$ \cite{fang2008scalable} & $20$ $\Omega$ \\
$D$ \cite{fang2008scalable} & $2.5*10^{-10} m^2$/s \\
$\delta$ \cite{fang2008scalable} & $25*10^{-9}$ m \\
$\kappa$ \cite{fang2008nonlinear} & $4*10^{-10} m^3$/C\\
		\bottomrule
	\end{tabular}
	\label{tab:2}
\end{table}

\subsection{Blocked force and free strain of FTCP actuator}
\begin{figure}[H]
      \centering
      {\includegraphics [trim=0cm 0.1cm 1cm 0.5cm, clip=true,width=0.5\linewidth]{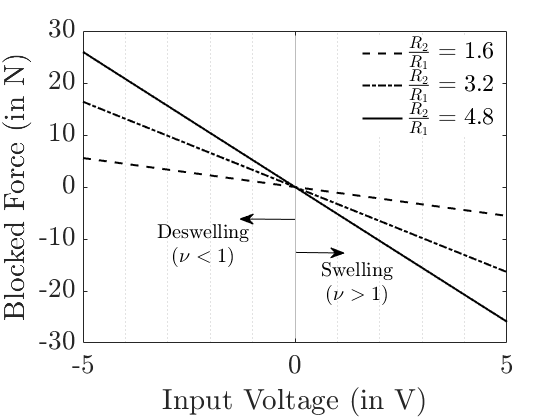}}
      \caption{Blocked force of FTCP actuator at applied voltage for different radius ratio.}
      \label{fig:5}
\end{figure}

\begin{figure}[H]
      \centering
      {\includegraphics [trim=0cm 0.1cm 1.1cm 0.5cm, clip=true,width=0.5\linewidth]{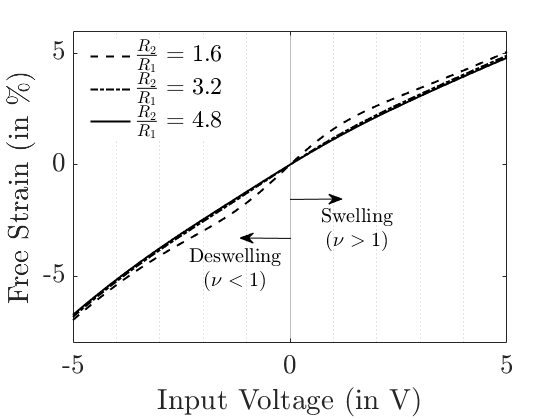}}
      \caption{Free expansion/contraction of FTCP actuator at applied voltage for various radius ratio.}
      \label{fig:6}
\end{figure}

The blocked force for an applied voltage can be obtained using equation \ref{eqn:34} at $\lambda_3 = 1$. The initial value of double layer (DL) capacitance in the simulation is chosen as $1.4*10^{-4} F$ for all results \citep{fang2008nonlinear}. Herein, we consider that the DL capacitance varies with a change in the surface area of the actuator; as the simulation progresses, as shown in Figure \ref{fig:2}. Figure \ref{fig:5} represents the blocked force versus applied voltage plots for different radius ratios. The radius ratio is defined as the outer radius to the inner radius of the actuator at reference configuration. The positive voltage swells the actuator resulting in a compressed load acting on one end of the actuator. On reversing the polarity of the applied voltage, the actuator shrinks to exert force in the opposite direction. The blocked force increases with an increase in radius ratio due to an increase in steady-state stored charge. Similarly, Figure \ref{fig:6} describes the free expansion/contraction of the FTCP actuator at different applied voltages for various radius ratios. The swelling ($\nu>1$) and deswelling ($\nu<1)$ effect results in axial elongation and contraction of the FTCP actuator, respectively. However, the free strain does not change with the radius ratio of the actuator, as shown in Figure \ref{fig:6}. This may be due to the assumption that DL capacitance varies as the actuator is stimulated. Consequently, the steady-state charge per unit of undeformed volume seems unaffected by an increase in the radius ratio of the actuator.

\subsection{Effect of double layer thickness on FTCP actuator}
\begin{figure}[H]
\centering
\begin{subfigure}{.49\textwidth}
  \centering
  {\includegraphics[trim=0cm 0.1cm 1cm 0.4cm, clip=true, width=.75\linewidth]{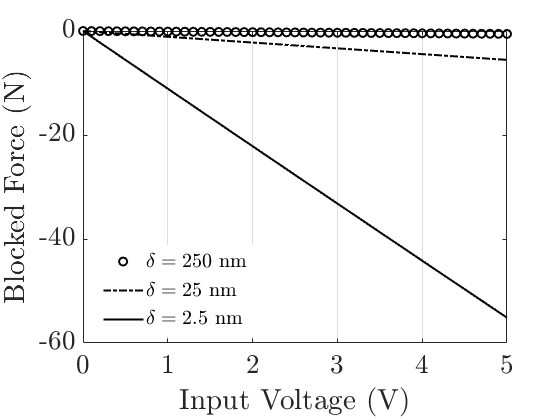}}
  \caption{}
  \label{fig:7sub1}
\end{subfigure}
\begin{subfigure}{.49\textwidth}
  \centering
  {\includegraphics[trim=0cm 0.1cm 1cm 0.4cm, clip=true,width=.75\linewidth]{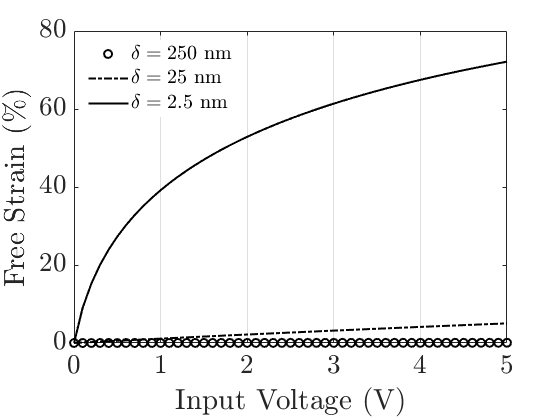}}
  \caption{}
  \label{fig:7sub2}
\end{subfigure}
     \caption{Effect of double layer thickness on (a) blocked force, and (b) free strain of FTCP actuator.}
      \label{fig:7}
\end{figure}

The steady-state charge stored in the FTCP actuator depends on DL thickness, as evident from equation \ref{eqn:117}. Figure \ref{fig:7} is drawn to see the effect of DL thickness on the actuator for the swelling case. The plot shows that both blocked force and free strain increases significantly with a decrease in double layer thickness. The possible reason for this behavior may be that the double layer thickness is inversely proportional to the ion concentration in bulk \citep{rossi2017modeling}. As a result, more number ions are available to diffuse, which causes a higher blocked force or free strain at a given voltage. Further, at $2.5$ nm DL thickness for the deswelling ($\nu < 1$) case shows erroneous output at higher voltages since the volume ratio of the actuator is becoming negative, which is non-physical. Thus, we suggest designing the deswelling actuator so that the value remains in the range of $25$ nm as given in other literature \citep{madden2000conducting,fang2008nonlinear,fang2010fiber}. Another possible way to utilize deswelling could be to operate it at a low voltage within $-1$V or tuning the fiber properties of the actuator as discussed in next subsection.

\subsection{Effect of fiber properties on FTCP actuator} 
\begin{figure}[H]
\centering
\begin{subfigure}{.49\textwidth}
  \centering
  {\includegraphics[trim=0cm 0cm 1.2cm 0.4cm, clip=true, width=0.75\linewidth]{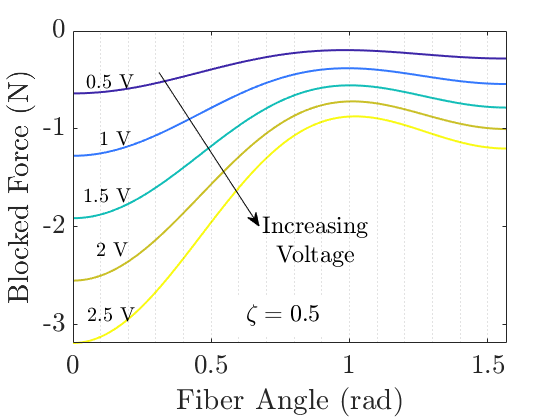}}
\end{subfigure}
\begin{subfigure}{.49\textwidth}
  \centering
  {\includegraphics[trim=0cm 0cm 0.9cm 0.6cm, clip=true,width=0.75\linewidth]{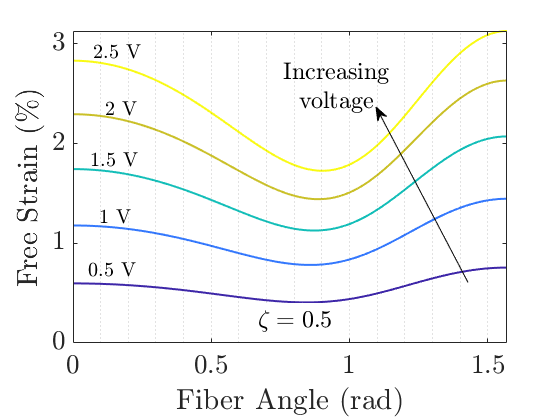}}
\end{subfigure}
\begin{subfigure}{.49\textwidth}
  \centering
  {\includegraphics[trim=0cm 0cm 1.2cm 0.4cm, clip=true,width=0.75\linewidth]{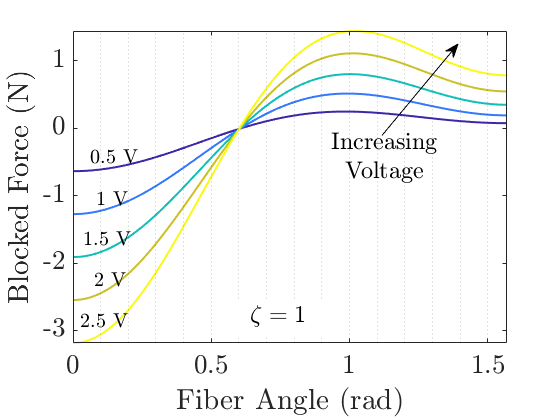}}
\end{subfigure}
\begin{subfigure}{.49\textwidth}
  \centering
  {\includegraphics[trim=0cm 0cm 0.9cm 0.4cm, clip=true,width=0.75\linewidth]{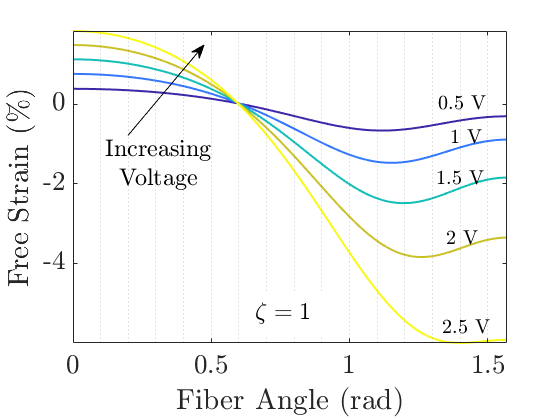}}
\end{subfigure}
\caption{Effect of fiber angle on FTCP actuator at various applied step voltage for two different fiber strength.}
\label{fig:8}
\end{figure}

Figure \ref{fig:8} shows the variation in the blocked force and free strain with fiber angle at different applied voltages for two anisotropy factors, i.e., $\zeta = 0.5$ and $\zeta = 1$. The applied voltage induces a constant swelling in the polymer. It is interesting to note that both blocked force and free strain change their behavior at higher anisotropy factor. In other words, at $\zeta = 1$, the blocked force change from compressive to tensile nature, and the free strain changes from elongation to contraction mode for same positive applied voltage at 0.6 radian ($\sim 34^0$) fiber angle. The model suggests that two different actuation modes (elongation and contraction) can be achieved with the same applied voltage by tuning the fiber angle.

\begin{figure}[H]
\centering
\begin{subfigure}{.49\textwidth}
  \centering
  {\includegraphics[trim=0cm 0cm 1cm 0.5cm, clip=true,width=.75\linewidth]{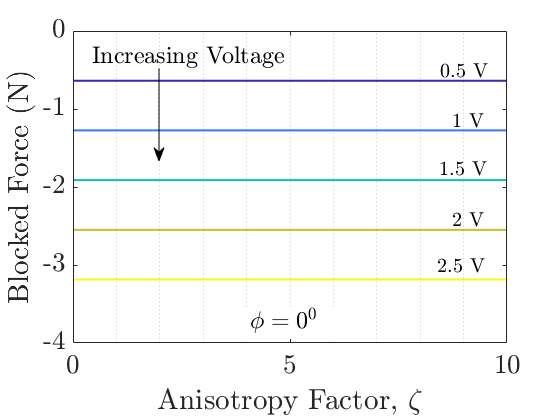}}
\end{subfigure}
\begin{subfigure}{.49\textwidth}
  \centering
  {\includegraphics[trim=0.2cm 0cm 1cm 0.5cm, clip=true, width=.75\linewidth]{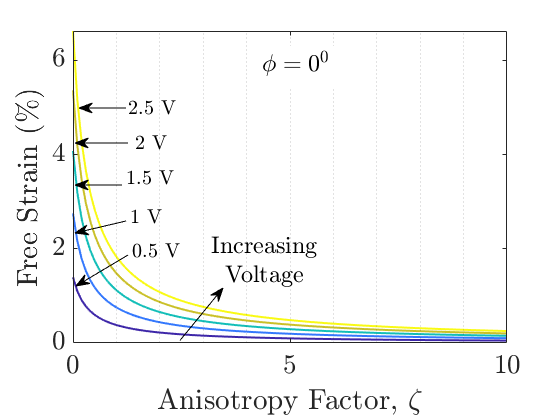}}
\end{subfigure}
\begin{subfigure}{.49\textwidth}
  \centering
  {\includegraphics[trim=0cm 0cm 1cm 0.5cm, clip=true,width=.75\linewidth]{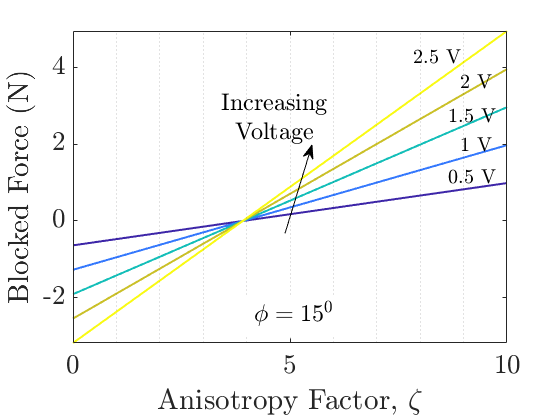}}
\end{subfigure}
\begin{subfigure}{.49\textwidth}
  \centering
  {\includegraphics[trim=0.2cm 0cm 1cm 0.5cm, clip=true,width=.75\linewidth]{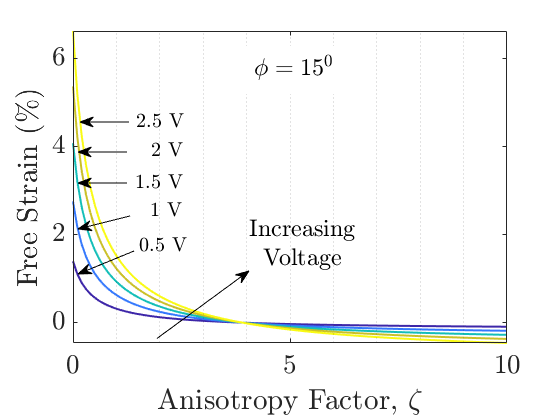}}
\end{subfigure}
\begin{subfigure}{.49\textwidth}
  \centering
  {\includegraphics[trim=0cm 0cm 1cm 0.5cm, clip=true,width=.75\linewidth]{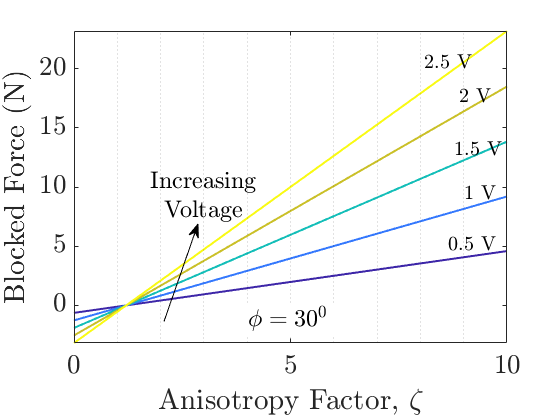}}
\end{subfigure}
\begin{subfigure}{.49\textwidth}
  \centering
  {\includegraphics[trim=0cm 0cm 1cm 0.5cm, clip=true,width=.75\linewidth]{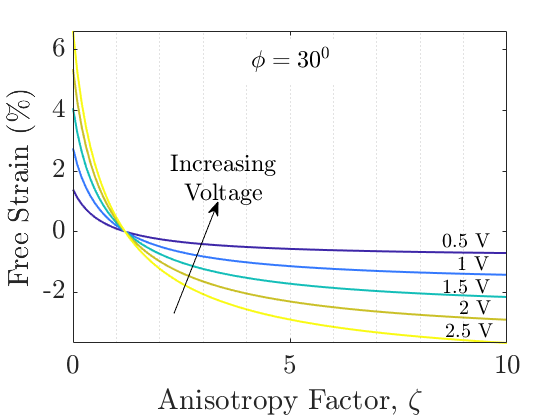}}
\end{subfigure}
\caption{Effect of anisotropy factor on FTCP actuator at various applied step voltage for different fiber angle.}
\label{fig:9}
\end{figure}

Figure \ref{fig:9} shows the change in blocked force and free strain of the actuator with anisotropy factor at various step voltages for varied fiber angle, i.e., $\phi = 0^0$, $\phi = 15^0$, and $\phi = 30^0$. The free strain varies nonlinearly with the model parameter, $\zeta$ at an input voltage for zero fiber angle. However, the blocked force shows negligible variation with anisotropy factor at an applied voltage for zero fiber angle. Further, it is interesting to observe here also we observe the same trend that axial elongation changes to contraction on increase of $\zeta$ for higher fiber angle ($\phi > 0^0$). Thus, FTCP actuators capable of axial elongation or contraction can be designed by adjusting the fiber characteristics. Furthermore, the different FTCP actuators can be combined to design soft locomotive robots \citep{shepherd2011multigait}. The design of a mechanically programmed low voltage-driven FTCP actuator similar to \citep{connolly2015mechanical} is the subject of future research.

\subsection{Experimental validation}
\subsubsection{Frequency response}
\begin{figure}[H]
      \centering
      {\includegraphics [trim=1.5cm 0cm 0.7cm 0.2cm, clip=true,width=0.45 \linewidth]{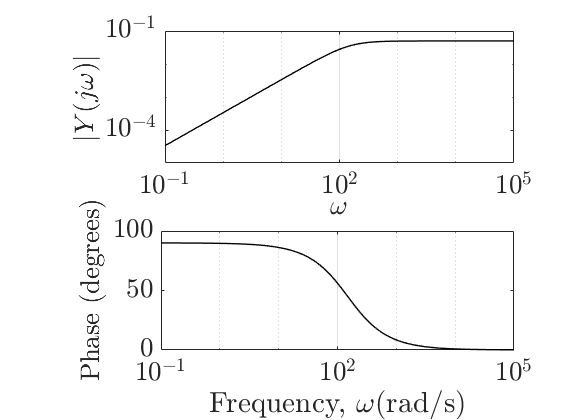}}
      \caption{Frequency response of the FTCP actuator.}
      \label{fig:11}
\end{figure}

The electrochemical process of the actuator is analyzed using an equivalent electrical circuit, as shown in Figure \ref{fig:3}. Admittance of the given circuit is estimated using (\ref{eqn:16}). Figure \ref{fig:11} shows the magnitude and phase angle of the circuit admittance. Since a similar circuit analogy is used here to describe the electrochemical process, the results available in the literature \citep{madden2000conducting, fang2008nonlinear} for different conducting polymer actuators show qualitatively similar plots, as shown in Figure \ref{fig:11}.  In addition, the resistance in the circuit dominates at high frequencies similar to \citep{madden2000conducting, fang2008nonlinear}. The magnitude of the admittance, $|Y(j\omega)| \longrightarrow \frac{1}{R_e}$ as $j\omega \longrightarrow \infty$. Also, the phase angle change shows that the circuit behavior shifts from capacitive to resistive as frequency changes from low to high. 
However, the stress-strain plot could be validated with experimental results for a special case of tubular conducting polymer actuator without fiber-reinforced.

\subsubsection{Stress strain behavior of the actuator}
\begin{figure}[H]
      \centering
      {\includegraphics [trim=0.2cm 0cm 1.2cm 0.4cm, clip=true,width=0.5\linewidth]{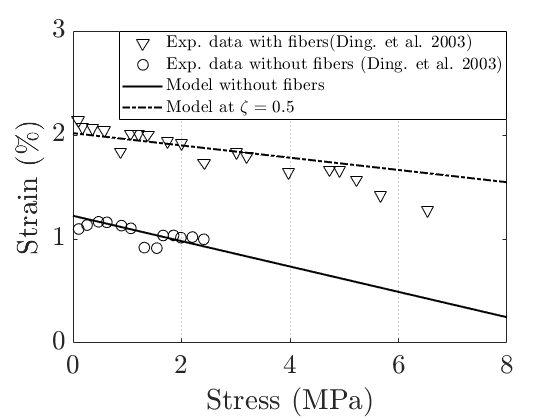}}
      \caption{Comparison of developed model (\ref{eqn:34}) without fibers and with fiber-reinforced ($\zeta = 0.5$) with  experimental data \citep{ding2003high} at 2.5 V.}
      \label{fig:12}
\end{figure}

We compared our model with available experimental data \citep{ding2003high} for tubular conducting polymer without fibers, i.e., $\zeta = 0$, and with fibers. The dimensions of the actuator, as listed in Table \ref{tab:2}, are consistent with the tubular CP actuator used in \citep{ding2003high}. Figure \ref{fig:12} presents the strain of the actuator for different isotonic stress levels at $2.5$ V. To compare the model with experimental data for tubular CP actuator without fibers, $\kappa = 10^{-10} m^3/C$ is considered  similar to \citep{madden2000conducting}. The model matches closely with experimental data for tubular CP actuators without fibers. Further, for FTCP actuator, $\zeta = 0.5$, and $\phi = 12^0$ \citep{ding2003high} is considered. A slight deviation from experiments can be found at higher applied loads for the FTCP actuator. As model considers a linear electro-chemo-mechanical coupling, which leads to a slight deviation from experiments for the FTCP actuator at higher applied loads \citep{sendai2009anisotropic}.

\section{Conclusion}\label{section-4}
This paper presents the electro-chemo-mechanical (ECM) deformation model of a fiber-reinforced tubular conducting polymer (FTCP) actuator. The electrochemical (EC) model is developed following an electrical circuit analogy that predicts the charge diffused inside FTCP for an applied voltage. The EC model (\ref{eqn:16}) of the actuator is expressed in terms of modified Bessel functions and circuit parameters. The volume change due to the ingress of ions is translated to axial elongation/contraction of the FTCP actuator using finite deformation theory. The developed model predicts the blocked force and free expansion/contraction of the actuator for an applied voltage.
Further, the model is used to analyze the effect of the electrical circuit and geometrical parameters on the actuation. An increase in output of the actuator is obtained with a decrease in double-layer thickness. Besides, the actuator shows dual behavior, i.e., axial elongation or contraction, depending upon the fiber properties at the same applied voltage. Thus, a combination of such actuators can lead to exciting applications in soft robotics. Furthermore, the frequency response plots are similar to previous literature \citep{madden2000conducting,fang2008nonlinear} which validates our EC model. The comparison of the developed ECM model with existing experimental results \citep{ding2003high} looks quite promising. Future research will involve detail experimental characterization of the FTCP actuator.

\section*{Acknowledgements}
The authors greatly acknowledge the computing facilities provided by the Indian Institute of Technology Delhi, India.

\section*{Declaration of competing interest}
The author(s) declared no potential conflicts of interest with
respect to the research, authorship, and/or publication of this
article.



\end{document}